# On the Performance of Reduced-Complexity Transmit/Receive-Diversity Systems over MIMO-V2V Channel Model

Yahia Alghorani and Mehdi Sayfi, *Members, IEEE*

*Abstract*—In this letter, we investigate the performance of multiple-input multiple-output techniques in a vehicle-to-vehicle communication system. We consider both transmit antenna selection with maximal-ratio combining and transmit antenna selection with selection combining. The channel propagation model between two vehicles is represented as $n*$Rayleigh distribution, which has been shown to be a realistic model for vehicle-to-vehicle communication scenarios. We derive tight analytical expressions for the outage probability and amount of fading of the post-processing signal-to-noise ratio.

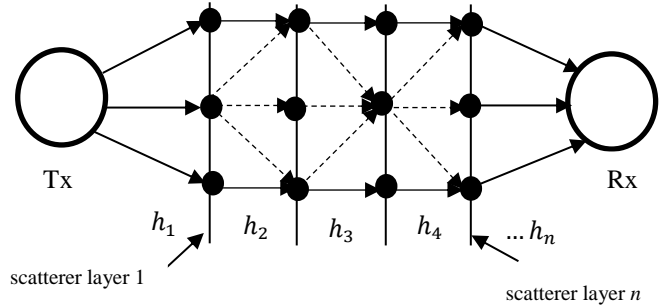

**Fig.1.** Multiple scattering model

## I. INTRODUCTION

Vehicle-to-vehicle (V2V) communication systems have received considerable attention lately due to the urgent need to develop a new road safety strategy. Future developments are expected in this field by using multiple antennas (MIMO) at the transmit and receive ends to enhance channel capacity and diversity (reliability), since multiple antenna elements can easily be placed on the large vehicle surface. To exploit the maximum transmit and receive diversity of MIMO systems, a transmit antenna selection with maximal-ratio combining (TAS/MRC) scheme was proposed in [1], as an attractive solution to provide a power efficient and low computational complexity, while maintaining the full diversity gain of conventional MIMO systems [2]. The main advantage of the TAS/MRC scheme is to reduce hardware cost due to the RF chains. By selecting only one transmit antenna that provides the highest post-processing signal-to-noise ratio (SNR), the feedback overhead of channel state information at the transmitter is significantly reduced compared to conventional MIMO systems. In this case, a single RF chain can be used at the transmit side regardless of the number of transmit antennas. To further reduce the number of expensive RF chains at the receive side, the transmit antenna selection with selection combining (TAS/SC) is implemented to select a single transmit and receive antenna, in which a single RF chain is used at both sides. Recently, a $n*$Rayleigh distribution was proposed as an accurate statistical channel model for mobile-to-mobile scenarios where both the transmitter and receiver are in motion and typically use the same antenna height, resulting in two or more small-scale processes generated by independent groups of scatterers around the two mobile terminals [3] (see Fig.1, where multiple scattering is taken place between the transmitter and the receiver, and all propagation paths travel through the same narrow pipe called by keyhole channels[1]). Such stochastic processes are widely encountered in dense urban and forest environments where local scattering objects such as buildings, street corners, signs, tunnels, bridges, moving vehicles, trees, and mountains, obstruct a direct radio wave path between the transmitter and the receiver giving rise to non-line-of-sight (NLOS) propagation. In other words, the $n*$Rayleigh fading channel can be defined as a product of $n$ independent Rayleigh random variables connected via narrow pipes and its probability density function (PDF) is given in [5]. Recent studies have proved that $n*$Rayleigh fading is an appropriate keyhole channel model for V2V communication [6]. For example, in [7], a new analytical expression for the joint PDF of bivariate double-Rayleigh distribution has been presented. In this study, the impact of outdated channel state information on a multichannel system has been reported. In [8], the authors studied the influence of interference sources in inter-vehicular communication systems (where both MRC reception and selection diversity are implemented over double-Rayleigh fading channels).

To the best of our knowledge, MIMO systems with antenna selection approaches such as TAS/MRC and TAS/SC schemes over $n*$Rayleigh fading channels have not been studied yet. Hence, it is the aim of this work to fill this research gap.

## II. SYSTEM MODEL

We consider a MIMO-V2V communication system equipped with $n_T$ transmit and $n_R$ receive antennas and signaling over a product of $n$ independent circularly complex Gaussian random variables, each having a channel coefficient between the $i$-th transmit antenna and $j$-th receive antenna equivalent to $h_{ij} \triangleq$

Yahia Alghorani is with the School of Electrical and Computer Engineering, University of California, Davis, CA US (e-mail: yalghorani@ucdavis.edu).

Mehdi Seyfi is with the School of Engineering Science, Simon Fraser University, Burnaby, BC V5A 1S6 Canada (e-mail: mehdi_seyfi@sfu.ca).

---

[1] Keyhole channel is defined as a multiplier between two fading processes, resulting in a received signal amplitude of a product of two Rayleigh random variables, e.g, double Rayleigh fading ($n = 2$) [4].

$\prod_{k=1}^{n} h_{ij,k}$ with zero mean and channel variance $\sigma_{ij}^2$ for $i = 1, 2, \ldots, n_T$, and $j = 1, 2, \ldots, n_R$. Therefore, the magnitude of channel coefficient $|h_{ij}|$ follows a $n*$Rayleigh distribution. The input-output relation for such a system is characterized by the $n_R \times n_T$ channel transfer matrix $\mathbf{H}$ consisting of $h_{ij}$ channel coefficients and rank $r \leq \min(n_T, n_R)$. We further assume that all underlying channels are quasi-static, which can be justified for V2V communications scenarios in rush-hour traffic. In addition, we assume that the channel matrix $\mathbf{H}$ is known at both the transmitter and the receiver. Let the $(n_T \times 1)$ symbols transmitted vector at the time instant $t$ be $\mathbf{x} = [x_1, x_2, \ldots, x_{n_T}]^T$, and the $(n_R \times 1)$ received signal vector is $\mathbf{y} = [y_1, y_2, \ldots, y_{n_R}]^T$, where the superscript denoted by $(.)^T$, is the transpose operator. By writing the received signal in vector form, we have

$$\mathbf{y} = \sqrt{P}\mathbf{H}\mathbf{x} + \mathbf{w}, \tag{1}$$

where $P$ is the total transmit power of the signal, and $\mathbf{w} = [w_1, w_2, \ldots, w_{n_R}]^T$ is a circularly symmetric complex-Additive Gaussian noise (AWGN) vector with zero mean and variance $N_o$, denoted as $w_j \sim \mathcal{CN}(0, N_o)$.

*A. TAS/MRC scheme*

In this scheme, a single antenna out of $n_T$ transmit antennas is selected to maximize the post-processing SNR ($\gamma_p = \|\mathbf{H}^{\max}\|^2 (P/N_o)$) at the maximal-ratio combining output, where the received signal vector can be expressed as

$$\mathbf{y} = \sqrt{P}\mathbf{H}^{\max}x + \mathbf{w}, \tag{2}$$

where $x$ is the transmit symbol. The squared channel vector norm is $\|\mathbf{H}^{\max}\|^2 = \underset{i \in n_T}{\mathrm{argmax}} \|\mathbf{H}_i\|^2 = \sum_{j=1}^{n_R} |h_{ij}|^2$, where $\mathbf{H}_i \in \mathcal{C}^{n_R \times 1}$ is the complex channel vector of the input complex Gaussian random variables. The instantaneous SNR at the receiver-MRC output is expressed as $\gamma_i = \sum_{j=1}^{n_R} |h_{ij}|^2 (P/N_o)$. Now, we can sort the RVs $\gamma_i$ which follow $n*$Rayleigh distribution in a descending order denoted as $\gamma_{n_T} \geq \cdots \gamma_i \geq \cdots \geq \gamma_1$, such that a transmit antenna is selected for the transmission per the $n_T$-th order statistics, where $\gamma_{n_T} = \gamma_p$. So, assuming $\{h_{ij}\}$ are independent and identically distributed (i.i.d) random variables with zero mean and variance $\sigma^2$, and using the approximate PDF for $n*$Rayleigh distribution given as [9]

$$f_X(x) \approx \frac{\beta^m}{n\Gamma(m)} x^{\alpha-1} e^{-\beta x^{1/n}}, \quad x \geq 0 \tag{3}$$

where $\alpha = m/n$ and $\beta = 2m/\Omega \bar{x}^{1/n}$, the approximate PDF of instantaneous SNR $\gamma$ at the receiver-MRC can be expressed as [10, Eq.(2)]

$$f_\gamma(\gamma) \approx \frac{\beta_{MRC}^a}{n\Gamma(a)} \gamma^{\alpha_{MRC}-1} e^{-\beta_{MRC}\gamma^{1/n}}, \tag{4}$$

where $a = mn_R$, $\alpha_{MRC} = a/n$, $\beta_{MRC} = 2a/\Omega(n_R \bar{\gamma})^{1/n}$, and $\bar{\gamma} = P\sigma^2/N_o$, $m$ and $\Omega$ are the fading severity parameters of $n*$Rayleigh fading channels given by [9].

$$m = 0.6102n + 0.4263, \quad \Omega = 0.8808n^{-0.9661} + 1.12$$

From (4), we calculate the approximate cumulative density function (CDF) of instantaneous SNR, $\gamma$, with the help of fact defined in [11, Eq.(3.381.1)], to be

$$F_\gamma(\gamma) \approx \frac{\gamma(a, \beta_{MRC}\gamma^{1/n})}{\Gamma(a)}, \tag{5}$$

where $\gamma(\alpha, x)$ is the lower incomplete gamma function, defined in [11]. Consequently, we determine the approximate CDF of the post-processing SNR $\gamma_p$ which is equivalent to the CDF of the largest of order statistics $\gamma_{n_T}$ [12, Sec. 2.1], as

$$F_{\gamma_p}(\gamma) = \Pr\{\gamma_p \leq \gamma\} = \Pr\{\text{all } \gamma_i \leq \gamma\}$$

$$\approx \left(1 - \frac{\Gamma(a, \beta_{MRC}\gamma^{1/n})}{\Gamma(a)}\right)^{n_T}, \tag{6}$$

where $\Gamma(\alpha, x)$ is the upper incomplete gamma function, defined in [11].

*B. TAS/SC scheme*

In this scheme, a single antenna out of $n_T$ antennas at the transmitter and a single antenna out of $n_R$ antennas at the receiver are jointly selected to maximize the post-processing SNR $(\gamma_p = |h^{\max}|^2(P/N_o))$ at the receiver terminal, where the received signal can be expressed as

$$\mathbf{y} = \sqrt{P}h^{\max}x + \mathbf{w}, \tag{7}$$

where the squared magnitude of the best channel coefficient between two terminals is $|h^{\max}|^2 = \underset{i \in n_T, j \in n_R}{\mathrm{argmax}} |h_{ij}|^2$. The instantaneous SNR between the $i$-th transmit antenna and $j$-th receive antenna can be expressed as $\gamma_{ij} = |h_{ij}|^2 P/N_o$. In this case, the RVs $\gamma_{ij}$ re-arranged in a descending order denoted as $\gamma_{n_T n_R} \geq \cdots \gamma_{ij} \geq \cdots \geq \gamma_{11}$, in which a single pair of antennas corresponding to the $n_T n_R$-th order statistics is selected for transmission and reception. Hence, the approximate CDF of the post-processing SNR $\gamma_p$ is given as

$$F_{\gamma_p}(\gamma) \approx \left(1 - \frac{\Gamma(m, \beta_{SC}\gamma^{1/n})}{\Gamma(m)}\right)^N, \tag{8}$$

where $N = n_T n_R$, $\beta_{SC} = 2m/\Omega \bar{\gamma}^{1/n}$.

III. PERFORMANCE ANALYSIS

*A. Outage Probability*

The outage probability $(P_{out} \triangleq F_{\gamma_p}(\gamma_o))$ of a communication channel is defined as the probability of the post-processing SNR $\gamma_p$ falls below a certain threshold ($\gamma_o = 2^R - 1$), namely

$$P_{out} = \Pr(\gamma_p \leq \gamma_o), \tag{9}$$

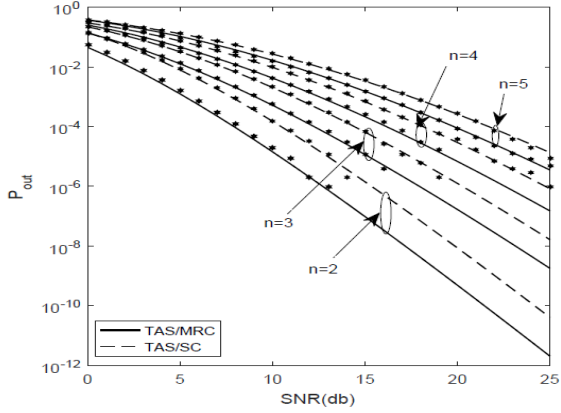

Fig.2. Comparison between the outage probability of TAS/MRC (6) and TAS/SC (8) schemes over $n$*Rayleigh fading channels. Monto-Carlo simulation is starred with antenna configuration: $n_T = 2$, $n_R = 3$, and $10^6$ iterations.

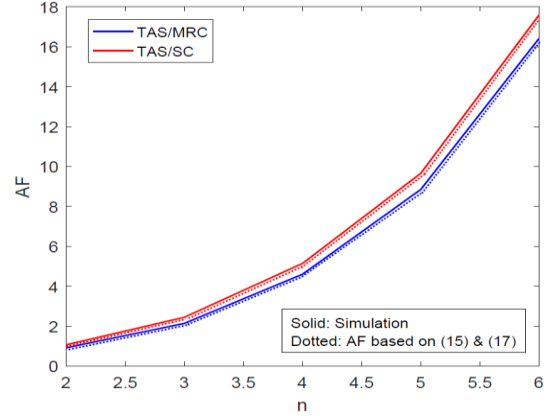

Fig.3. Amount of fading of TAS/MRC and TAS/SC schemes over $n$*Rayleigh fading channels ($n_T = 2$, $n_R = 2$). Weighting coefficients are set to ($b_1 = 2.3, b_2 = 1.5, n = 2$), ($b_1 = 2.1, b_2 = 1.5, n = 3$), ($b_1 = 2, b_2 = 1.5, n = 4$), ($b_1 = 1.57, b_2 = 1.5, n = 5$), and ($b_1 = 1.44, b_2 = 1.68, n = 6$).

where $R$ is a target transmission rate.

*1) TAS/MRC scheme*

Replacing (6) into (9), the approximate expression of the outage probability for TAS/MRC scheme is obtained. To compute the maximum achievable diversity order, we need to find an asymptotic expression for the outage probability in the high SNR regime (i.e., when $\bar{\gamma} \to \infty$), using the fact that [11, Eq.(8.354.2)], which leads to $\Gamma(\alpha, x) = \Gamma(\alpha) - x^\alpha/\alpha$ when $x \to 0$. Thus, we can rewrite (9) in an asymptotic form as

$$P_{out} \approx \left(\frac{(2\,a/\Omega)^a}{a\Gamma(a)}\right)^{n_T} z^d + \mathcal{O}(z^{d+1}), \quad (10)$$

where $z = \frac{\gamma_o}{n_R \bar{\gamma}}$. Subsequently, from (10), the achievable diversity order is calculated as $d = mN/n$, with a coding gain equivalent to

$$CG = \left(\frac{(a\Gamma(a))^{1/a}}{2\,a/\Omega\,n_R^{1/n}}\right)^n. \quad (11)$$

Note that $d \approx N$ when $n = 2$. Also, the coding gain in (11) depends on both the fading severity parameters and the number of receive antennas regardless the number of transmit antennas.

*2) TAS/SC scheme*

Replacing (8) into (9), the approximate expression of the outage probability for TAS/SC scheme is found. Thus, at the high SNR values, (9) can be rewritten in an asymptotic form as

$$P_{out} \approx \left(\frac{(m/\Omega)^m}{m\Gamma(m)}\right)^N z^d + \mathcal{O}(z^{d+1}). \quad (12)$$

From (10) and (12), both the TAS/MRC and TAS/SC schemes enjoy the same diversity order of $d = mN/n$, but TAS/SC scheme provides a coding gain equal to

$$CG = \left(\frac{(m\Gamma(m))^{1/m}}{2m/\Omega}\right)^n. \quad (13)$$

Note that the coding gain in (13) depends only on the fading severity parameters, which is assumed to be fixed during the whole transmission time regardless the number of transmit and receive antennas. Fig.2 compares the outage probability of TAS/MRC and TAS/SC schemes with antenna configuration $n_T = 2$, $n_R = 3$. We assume that the fading severity parameter, $n$, takes the values $n = 2, 3, 4$ and $5$, and Monte-Carlo simulations are performed to show the accuracy of (6) and (8). In our simulation, by testing the values of $n$ from 2 to 5, the parameter, $\beta_{MRC}$, is weighted by a factor $\omega = 1.176$ (i.e. $\omega\,\beta_{MRC}$). This is because of the approximation error in (6); whereas we keep the factor $\omega = 1$ for the case of TAS/SC (where the approximation error is minimized by selecting only one antenna at the receiver).

In general, from Fig.2, an exact match is evident between simulations and analytical results. The outage performance of both schemes deteriorates by increasing the fading severity parameter $n$, confirming our observations that the diversity order $(d = m\,N/n)$ decreases when $n$ increases. Plus, TAS/MRC enjoys better performance over cascaded Rayleigh fading channels. Based on these observations, it is important to consider the dynamic range of measurement devices for detecting such severe fading channels. For instance, when the outage probability based on (6) is assumed at $P_{out} = 10^{-4}$, the required minimum SNR levels for receiving an undistorted signal are 8, 13, 17.5, and 21.4 dB for $n = 2, 3, 4$, and 5, respectively. Therefore, antenna selection schemes can play a key role in enhancing a SISO system dynamic range [5].

*B. Amount of Fading*

In this subsection, we present approximate closed-form expression of the amount of fading (AF) for both TAS/MRC and TAS/SC schemes over $n$*Rayleigh fading channels. To evaluate the severity of fading at the receiver, we derive the approximate $l$-th moment of $\gamma_p$, $\mathbf{E}[\gamma_p^l]$, which is used to calculate the fading figure defined by [13]

$$AF = \frac{\mathbf{E}[\gamma_p^2]}{(\mathbf{E}[\gamma_p])^2} - 1, \quad (14)$$

where $\mathbf{E}[.]$ denotes the statistical average.

*1) TAS/MRC scheme*

In this scheme, the approximate *l*-th moment of $\gamma_p$ can be derived from (6) with the help of [11, Eq.(3.326.2)] and the bounds for the incomplete gamma function given in [14, Eq.(1.5)], to be finally expressed as

$$\mathbf{E}[\gamma_p^l] \approx \sum_{k=1}^{n_T} \binom{n_T}{k}(-1)^k \frac{b_1(a_k \Gamma(a_k + nl) - \Gamma(a_k + nl + 1))}{\Gamma(a) k^{a_k+nl} \beta_{MRC}^{nl}}, \quad (15)$$

where $a_k = k(a-1)$, $b_1 > 1$ is a weighting coefficient. By determining the first and the second moments of $\gamma_p$, we calculate the AF lower bound of the considered scheme. To gain further insight into the AF performance over cascaded Rayleigh fading channels, we can use the bound of the incomplete gamma function given in [15]

$$\text{AF} \approx \frac{(1+mN)^{mN}(a\Gamma(a))^{n_T}\Gamma(mN+2n)}{(a+1)^{mN} mN \Gamma^2(mN+n)} - 1.$$

By fixing the number of transmit antennas at $n_T = 2$, the AF lower bound expression is greatly reduced to $\text{AF} \approx 2^{n/n_R} - 1$.

*Special Case*: For the SIMO-MRC channel model, (14) can be rewritten in a simple compact form as

$$\text{AF}_{\text{SIMO}} \approx \frac{\Gamma(a)\Gamma(a+2n)}{\Gamma^2(a+n)} - 1. \quad (16)$$

For double-Rayleigh distribution ($n=2$), (16) can be simply expressed as $\text{AF}_{\text{SIMO}} \approx 2.5^{2/n_R} - 1$.

*2) TAS/SC scheme*

In this scheme, the approximate *l*-th moment of $\gamma_p$ is derived from (8) as

$$\mathbf{E}[\gamma_p^l] \approx \sum_{k=1}^{N} \binom{N}{k}(-1)^k \frac{b_2(a_k \Gamma(a_k + nl) - \Gamma(a_k + nl + 1))}{\Gamma(m) k^{a_k+nl} \beta_{SC}^{nl}} \quad (17)$$

where $a_k = k(m-1)$, $b_2 > 1$. By computing the first two moments, we calculate the approximate AF of the underlying scheme.

*Special Case*: For the SISO channel model, (14) can be rewritten in a compact form as

$$\text{AF}_{\text{SISO}} \approx \frac{\Gamma(m)\Gamma(m+2n)}{\Gamma^2(m+n)} - 1. \quad (18)$$

For classical Rayleigh distribution ($n=1$), the corresponding $\text{AF}_{\text{SISO}}$ is equivalent to $\text{AF}_{\text{SISO}} \approx 1$. From (16) and (18), we also obtain a trade-off expression $\text{AF}_{\text{SIMO}} \approx \text{AF}_{\text{SISO}}/n_R$ for $n=1$ and $n=2$. Our analytical results generally reveal that the fading figure over $n$*Rayleigh is more severe than that of classical Nakagami-*m* fading with $\text{AF} = 1/m$ and Ricean fading with $\text{AF} = (1+2K)/(1+K)^2$. Moreover, TAS/MRC scheme outperforms TAS/SC in reducing the amplitude of fading at the receiver; i.e., as the number of receive antennas increases, the amount of fading is reduced. In Fig.3, We illustrate the impact of transmit antenna selection schemes on cascaded Rayleigh fading channels. The tightness of the AF lower-bound based on (15) and (17), is verified by Monte-Carlo simulation. The figure shows that the TAS/MRC scheme outperforms the TAS/SC in combating the effect of $n$*Rayleigh fading channels and the performance difference between the two schemes becomes larger as $n$ increases.

## IV. CONCLUSION

In this letter, we investigate the performance of reduced-complexity transmit and receive diversity over V2V channels. In particular, the performance of TAS/MRC and TAS/SC schemes was analyzed over $n$*Rayleigh fading channels. Novel closed-form expressions were derived for the outage probability and the amount of fading. Our diversity order analysis shows that a maximum diversity order of $(mN/n)$ is achieved for both schemes. In addition, our numerical results demonstrated that TAS/MRC outperforms TAS/SC, but its performance is limited as $n$ increases. In summary, the $n$*Rayleigh channel fading model achieves a good cost-performance tradeoff for V2V communications systems when the number of RF chains is limited.

## REFERENCES

[1] S. Thoen, L. Van der Perre, B. Gyselinckx, and M. Engels, "Performance analysis of combined transmit-SC/receive-MRC," *IEEE Trans. Commun.*, vol. 49, pp. 5–8, Jan. 2001.
[2] A. F. Molisch, M. Z. Win, and J. H. Winters, "Reduced-complexity transmit/receive diversity systems," *IEEE Trans. Signal Process*, vol. 51, no. 11, pp. 2729–2738, Nov. 2003.
[3] I. Z. Kovacs, "Radio channel characterisation for private mobile radio systems: Mobile-to-mobile radio link investigations," Ph.D.dissertation, Aalborg Univ., Aalborg, Denmark, Sep. 2002.
[4] D. Chizhik, G. Foschini, M. Gans, and R. Valenzuela, "Keyholes, correlations and capacities of multielement transmit and receive antennas," *IEEE Trans. Wireless Commun.*, vol. 1, pp. 361–368, Apr. 2002.
[5] J. Salo, H. E. Sallabi, and P. Vainikainen, "The distribution of the product of independent Rayleigh random variables," *IEEE Trans. Antennas Propagat.*, vol. 54, no. 2, pp. 639–643, Feb. 2006.
[6] D. W. Matolak and J. Frolik, "Worse-than-Rayleigh fading: Experimental results and theoretical models," *IEEE Commun. Mag.*, vol. 49, no. 4, pp. 140-146, Apr. 2011.
[7] P. S. Bithas, K. Maliatsos and A. G. Kanatas, "The bivariate double Rayleigh distribution for multichannel time-varying systems," *IEEE Wireless Commun Letters*, vol. 5, no. 5, pp. 524-527, Oct. 2016.
[8] P. S. Bithas, G. P. Efthymoglou and A. G. Kanatas, "Intervehicular communication systems under co-channel interference and outdated channel estimates," *2016 IEEE International Conference on Communications (ICC)*, Kuala Lumpur, pp. 1-6, 2016.
[9] H. Lu, Y. Chen, and N. Cao, "Accurate approximation to the PDF of the product of independent Rayleigh random variables," *IEEE Antennas Wireless Propag. Lett.*, vol. 10, pp. 1019–1022, Oct. 2011.
[10] Y. Alghorani, G. Kaddoum, S. Muhaidat, S. Pierre and N. Al-Dhahir, "On the Performance of Multihop-Intervehicular Communications Systems Over n*Rayleigh Fading Channels," *IEEE Wireless Commun Letters*, vol. 5, no. 2, pp. 116-119, April 2016.
[11] I. S. Gradshteyn and I. M. Ryzhik, *Table of Integrals, Series and Prod Ucts*, 7th ed. New York: Elsevier, 2007.
[12] H. A. David, *Order Statistics*. New York: Wiley, 1970.
[13] M. K. Simon and M.-S. Alouini, *Digital Communication Over Fading Channels*, 2nd ed. New York: Wiley, 2004.
[14] P. Natalini and B. Palumbo, "Inequalities for the incomplete gamma function," Mathematical Inequalities & Applications, vol. 3, no. 1, pp. 69–77, 2000.
[15] E. Neuman, Inequalities and Bounds for the Incomplete Gamma Function, *Results. Math.* 63 (2013), 1209–1214.